\documentclass[10pt, conference, compsocconf]{IEEEtran}
\pdfoutput=1 
\IEEEoverridecommandlockouts
\usepackage{cite}
\usepackage{amsmath,amssymb,amsfonts}
\usepackage{graphicx}
\usepackage{textcomp}
\usepackage{xcolor}
\def\BibTeX{{\rm B\kern-.05em{\sc i\kern-.025em b}\kern-.08em
    T\kern-.1667em\lower.7ex\hbox{E}\kern-.125emX}}

\usepackage[hidelinks,
pdfauthor={Xuewei Ma, Geng Qin, Zhiyang Qiu, Mingxin Zheng, Zhe Wang},
            pdftitle={RiWalk: Fast Structural Node Embedding via Role Identification},
            pdfsubject={},
            pdfkeywords={structural embedding, network embedding, graph kernel, structural role}
            ]{hyperref}

\usepackage{algorithm}
\usepackage{algcompatible}
\renewcommand{\COMMENT}[2][.5\linewidth]{%
  \leavevmode\hfill\makebox[#1][l]{//~#2}}
\algnewcommand\algorithmicto{\textbf{to}}
\algnewcommand\RETURN{\State \textbf{return} }

\usepackage{url}
\usepackage{booktabs}
\usepackage{xspace}

\usepackage{multirow}
\usepackage{subfigure}
\usepackage[inline]{enumitem}

\def\ourwork{RiWalk\xspace}
\def\structovec{\textit{struc2vec}\xspace}
\def\nodetovec{\textit{node2vec}\xspace}
\def\graphwave{\textit{GraphWave}\xspace}
\def\deepwalk{\textit{DeepWalk}\xspace}
\def\wordtovec{\textit{word2vec}\xspace}
\def\LINE{\textit{LINE}\xspace}
\def\rolx{\textit{RolX}\xspace}

\newcommand{\Graph}{G}
\newcommand{\Vertices}{V}
\newcommand{\Edges}{E}

\newcommand{\Nei}{\mathcal{N}}
\newcommand{\nx}{\mathbf{x}}

\newcommand{\eg}{\emph{e.g.}}
\newcommand{\ie}{\emph{i.e.}}

\newcommand{\wrt}{\emph{w.r.t.}\ }

\begin{document}

\title{\ourwork: Fast Structural Node Embedding via\\ Role Identification}

\author{
\IEEEauthorblockN{Xuewei Ma\IEEEauthorrefmark{2}\IEEEauthorrefmark{3}, Geng Qin\IEEEauthorrefmark{2}\IEEEauthorrefmark{3}, Zhiyang Qiu\IEEEauthorrefmark{2}, Mingxin Zheng\IEEEauthorrefmark{2}\IEEEauthorrefmark{3}, Zhe Wang\IEEEauthorrefmark{2}\IEEEauthorrefmark{3}\thanks{\IEEEauthorrefmark{1}Corresponding author.}\IEEEauthorrefmark{1}}
\IEEEauthorblockA{\IEEEauthorrefmark{2}College of Computer Science and Technology, Jilin University, Changchun, 130012, China.}
\IEEEauthorblockA{\IEEEauthorrefmark{3}Key Laboratory of Symbol Computation and Knowledge Engineering, Ministry of Education,\\ Jilin University, Changchun, 130012, China.}

xuew.ma@gmail.com; \{qingeng17, qiuzy2118, zhengmx17\}@mails.jlu.edu.cn; wz2000@jlu.edu.cn
}

\maketitle

\begin{abstract}
Nodes performing different functions in a network have different roles, and these roles can be gleaned from the structure of the network. 
Learning latent representations for the roles of nodes helps to understand the network and to transfer knowledge across networks.
However, most existing structural embedding approaches suffer from high computation and space cost or rely on heuristic feature engineering. 

Here we propose \ourwork, a flexible paradigm for learning structural node representations. It decouples the structural embedding problem into a role identification procedure and a network embedding procedure. Through role identification, rooted kernels with structural dependencies kept are built to better integrate network embedding methods. 
To demonstrate the effectiveness of \ourwork, we develop two different role identification methods named \ourwork-SP and \ourwork-WL respectively and employ random walk based network embedding methods. 

Experiments on within-network classification tasks show that our proposed algorithms achieve comparable performance with other baselines while being an order of magnitude more efficient.
Besides, we also conduct across-network role classification tasks. The results show potential of structural embeddings in transfer learning.
\ourwork is also scalable, making it capable of capturing structural roles in massive networks. 
\end{abstract}


%\IEEEpeerreviewmaketitle

\section{Introduction}
Nodes in the same network always perform different functions and have different behaviors, leading them to different roles, \eg, leaders of communities or bridges between groups.  
At the same time, nodes of different networks may share similar roles, \eg, hubs of airline networks and managers of companies.
Identifying roles in networks will help researchers to gain a thorough understanding of network evolution and to transfer their knowledge across networks, thus leading to better performances of several real-world tasks, which may include fraud detection~\cite{akoglu2013opinion}, network integration~\cite{heimann2018regal}, protein function prediction of protein interaction networks~\cite{milenkovic2008uncovering} or individualized recommendation~\cite{pirotte2007random}.

\begin{figure}[t]
  \centering
  \includegraphics[width=\linewidth]{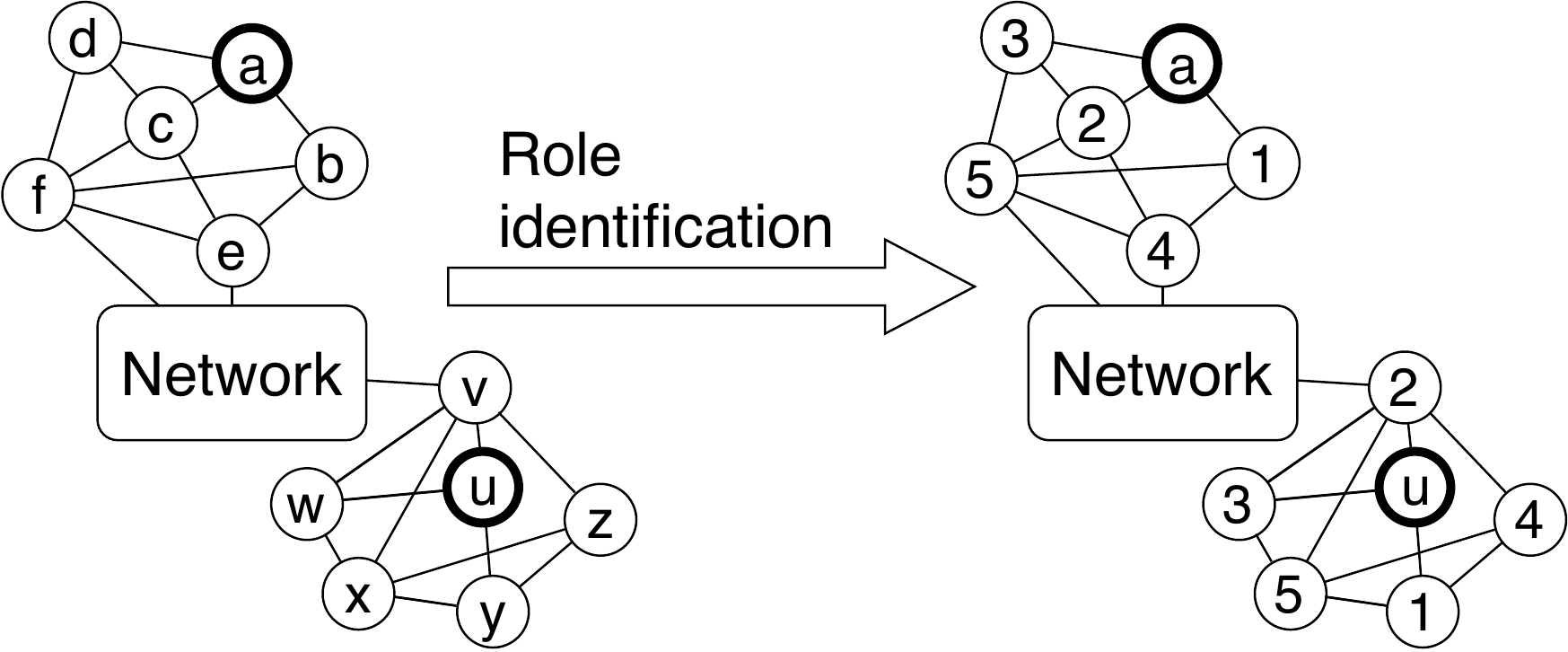}
  \caption{Two nodes $a$ and $u$ residing far apart in a network have similar local topologies but totally different context nodes. 
However, after the role identification procedure, they have similar context and are indirectly densely connected, thus typical network embedding methods can be directly applied to learn structural embeddings.}
\label{fig:role-identification}
\end{figure}

Intuitively, roles can be reflected in the network topologies, for example, leaders are corresponding to hub nodes of cliques and bridges between groups are corresponding to structural holes. 
In fact, structural role identification has been studied for decades. However, despite their achievements, most traditional approaches~\cite{leicht2006vertex,zager2008graph,jin2011axiomatic} require computation of multiple complicated graph-metrics, such as PageRank value~\cite{page1999pagerank}, structural hole value~\cite{burt2009structural} or clustering coefficient~\cite{watts1998collective}. 
Thus, these approaches are often ad hoc and time-consuming, making them hard to extend to massive networks.

On the other hand, graph representation learning (also known as network embedding), recently shows great potential for capturing neighborhood similarities and community memberships in low-dimensional representations. 
The learned embeddings can be used as features to boost the results of downstream machine learning tasks. 
Due to its simplicity, efficiency, and scalability, graph representation learning makes it possible for machine learning on large-scale networks.

However, most state-of-the-art network embedding algorithms~\cite{grover2016node2vec,perozzi2014deepwalk,tang2015line} assume that the more two nodes are densely connected, the more similar they are supposed to be. 
Thus, two nodes residing distantly in a network will be embedded far apart, even if they may share similar roles in the network. 
Therefore, it is impossible to directly apply typical network embedding methods to learn structural node representations. 

Several approaches~\cite{donnat2018learning,ribeiro2017struc2vec} have been proposed to fill this gap in recent years. 
Despite their success, these methods do not make full use of existing network embedding approaches, making them either rely heavily on heuristic feature engineering or suffer from high computation and space cost, thus hard to generalize across graphs and scale to massive networks. 
More details about existing methods will be discussed in Section \ref{sec:related-work}.

To overcome the above limitations, we propose \ourwork (\textit{Ri} as \textit{role identification}), an algorithmic paradigm that relabels context nodes with their structural roles in the anchor nodes' local subgraphs (as illustrated in Fig.~\ref{fig:role-identification}), and thus typical network embedding methods can be directly applied to learn structural embeddings.

The key ideas of \ourwork are as follows: 
\begin{itemize}
\item \ourwork decouples the structural embedding problem into two procedures: the role identification procedure and the network embedding procedure. 

\item The role identification procedure takes the idea of substructure and relabeling from graph kernel methods. However, distinct from traditional graph kernels, structural dependencies between substructures are kept after role identification.
Through role identification, graph kernel methods and network embedding algorithms are better integrated compared to prior works, thus leading to a lower complexity.

\item Since \ourwork is a general meta-strategy, both the role identification procedure and the embedding procedure can be customized for particular purposes.
The framework is simple and efficient because of the decoupling, making it easy to apply to different graphs.
\end{itemize}

To demonstrate the effectiveness of our proposed model, we put forward two different role identification methods named \ourwork-SP and \ourwork-WL,  respectively. 
\ourwork-SP is based on an intuition that two nodes are structurally similar if their neighbor nodes have similar degrees. The identifiers generated by \ourwork-SP are in line with the substructures of the Shortest-Path graph kernel.
\ourwork-WL takes the idea of neighborhood aggregation from the Weisfeiler-Lehman graph kernel.
It considers permutations of distances to signify the relative positions of context nodes from the anchor node, resulting in an ability to capture more fine-grained connectivity patterns.

We compare our proposed methods with state-of-the-art baselines~\cite{grover2016node2vec,donnat2018learning,ribeiro2017struc2vec} on several real-world datasets. 
Firstly, we illustratively demonstrate the difference between typical network embededing and structural embedding on an expressway network. 
Results show that typical network embedding methods assume the predicted labels to be smoothly distributed over the network. Thus they are not capable of identifying roles of nodes, since roles always distribute non-smoothly over graphs.

Then we carry out two node classification tasks: a within-network node classification task, where the training data and the testing data come from the same network and an across-network classification task, where classifiers are trained on one network but used to predict the class labels of the nodes in other networks. Besides, an experiment of structural hole identification is also conducted.
The results of the these experiments prove the effectiveness and efficiency of our proposed model.
Specifically,
while \ourwork-SP shares similar intuition with our baseline method, by using our proposed paradigm, \ourwork-SP learns more robust node representations. 
Meanwhile, we show that our proposed algorithms achieve comparable performances with other baselines while being an order of magnitude more efficient.

To summarize, our paper makes the following contributions:
\begin{enumerate}
\item We propose the \ourwork, a novel paradigm for learning structural embeddings of nodes. 
It is based on structural role identifications of subgraphs and can better integrate graph kernels with network embedding methods to leverage recent advancements in network embedding.
\item We put forward two different role identification methods
and demonstrate that despite its simplicity, this framework performs surprisingly well on real-world datasets and outperforms state-of-the-art baselines.
\item We discuss the gaps between typical network embedding and structural embedding and give an example to demonstrate the differences.
\item We conduct an across-network node classification task, which proves that structural embeddings can be used for transfer learning.
\end{enumerate}

The rest of this paper is arranged as follows. 
In Section~\ref{sec:related-work}, we briefly discuss related works along with their limitations. 
Section~\ref{sec:preliminaries} formally defines the problem of structural role embedding and briefly introduces two related graph kernel methods. 
We present the \ourwork\ framework in details in Section~\ref{sec:ourwork}. 
In Section~\ref{sec:experiments}, we empirically evaluate \ourwork on real-world graphs. 
Finally, we conclude in Section~\ref{sec:conclusion}.

\section{Related Work}
\label{sec:related-work}
Our work is inspired by two lines of research: 
\begin{enumerate*}
\item graph kernels and 
\item network embedding.
\end{enumerate*}
\paragraph{Graph Kernels} Graph kernels are functions that measure similarities between graphs. Most graph kernel methods~\cite{borgwardt2005shortest,shervashidze2009fast,shervashidze2009efficient} can be viewed as R-convolution kernels, which first decompose graphs into substructures and then compute graph similarities based on similarities between these components~\cite{vishwanathan2010graph}. With different definition of substructure or different choice of similarity measure, one can define different graph kernels.

\paragraph{Network Embedding} Network embedding methods aim to learn distributed representations of nodes in a network.
\deepwalk~\cite{perozzi2014deepwalk} first leverages \wordtovec~\cite{mikolov2013distributed} to learn a language model for a network by treating truncated random walks as sentences and neighbor nodes as semantically similar words. 
\nodetovec~\cite{grover2016node2vec} generalizes this model by adding flexibility in neighborhood exploring. 
\LINE~\cite{tang2015line} learns embeddings through optimizing an objective function which preserves both the first- and the second-order proximities.
Through these models, distributed representations can successfully capture neighborhood similarities and be directly used as features for machine learning tasks. 
\\

Arguably, most presented network embedding approaches learn representations of nodes by minimizing the distance between two distributions. Both the two distributions represent the conditional probabilities of context nodes ``generated'' by the anchor nodes but one is specified by the embeddings and the other one is empirical. 
For example, in \deepwalk and \nodetovec, context nodes are a set of nodes which appear in the same fixed-sized sliding window with the anchor node in random walks, and in \LINE, context nodes are directly linked with the anchor node (first-order proximity) or share the same neighbor node with the anchor node (second-order proximity). 
Therefore, two nodes will be embedded closely only if they are close and densely connected in the network. 
When we want two structurally similar nodes which reside distantly (or even in different isolated parts) to be embedded closely, typical network embedding approaches are not suitable.

Considerable effort has been made for learning structural embeddings of nodes recently, for example, \structovec~\cite{ribeiro2017struc2vec} and \graphwave~\cite{donnat2018learning}. 
\structovec takes three steps to get structural embeddings. First, it utilizes a simplified graph kernel to measure structural similarities between nodes in different neighborhood sizes. Secondly, it encodes these similarities into weights of edges to construct a hierarchy of complete graphs, such that edges between structurally similar nodes have small weights. Thirdly, a random walk based embedding method is used to learn node embeddings. 
While this method is effective, it suffers from high time and space cost due to the pair-wise subgraph comparison and complete graph construction.

\graphwave applies graph diffusion kernels~\cite{coifman2006diffusion} and treats these kernels as probability distributions over the graph. 
Then embeddings can be obtained by characterizing the distributions using empirical characteristic functions~\cite{lukacs1970characteristic}. 
Since it needs to compute a full eigendecomposition of a large matrix, \graphwave suffers high space cost. 

Note that though several approaches~\cite{ahmed2018learning,cao2016deep,wang2016structural} have been proposed to learn representations with structural information, these methods are not designed to learn pure topological embeddings. 
Since pure topological embeddings can be used for tranfer learning (to be show in \ref{sec:transfer}), while neighborhood-based embeddings are not, structural representation learning is somewhat orthogonal to typical network embedding.

\section{Preliminaries}
\label{sec:preliminaries}

\subsection{Problem Formulation}

Let $\Graph=(\Vertices,\Edges)$ denote an undirected unweighted graph, where $\Vertices$ is the set of $\lvert \Vertices \rvert$ nodes $\{v_1,v_2\ldots v_{\lvert \Vertices \rvert}\}$ and $\Edges \subseteq \Vertices \times \Vertices$ is the set of $\lvert \Edges \rvert$ edges. 
Given only $G$ itself, we aim to learn a mapping function from nodes to low-dimensional representations $f:\Vertices \to \mathbb{R}^{d}$, where $d \ll \lvert \Vertices \rvert$ is the number of dimensions of the learned representations. 
In the latent space, we hope that structural similar nodes are represented closely and nodes with different local structures are represented far apart.

Here we introduce the basic notations and terminologies which will be used throughout this paper. 
Given a graph $\Graph$ defined as above, we use $s_{ij}$ to denote the distance (\ie, the length of the shortest path) between two nodes $v_i$ and $v_j$.
Then we let $k^*$ to denote the diameter (\ie, the length of the longest shortest path betweeen any pairs of nodes) of $\Graph$. 
Given a specific node $v_i$, the set of neighbor nodes within radius $k$ of $v_i$ is defined as $\Nei_i^k=\{v_j \in \Vertices \mid s_{ij}\leq k\}$.
Thus for each $v_i \in \Vertices$, $\Vertices \equiv \Nei_i^{k^*}$.
Letting $\Nei_i=\Nei_i^1$ denote the set of direct neighbor nodes of $v_i$, we use $\delta_i=\lvert \Nei_i \rvert$ to denote its degree. 
Given a set of nodes $S\subseteq \Vertices$, we use $\Graph(S)$ to denote the subgraph induced by $S$.

\subsection{Graph Kernels}
Firstly, we summarize the core concepts of graph kernels and introduce the two of them that are related to ourwork. Given a set of graphs $\mathcal{G}$, R-convolution kernels first decompose each graph into atomic substructures. Then for each two graphs $\Graph, \Graph' \in \mathcal{G}$, a kernel function $\mathcal{K}$ is defined as 
\begin{equation}
\mathcal{K}(\Graph,\Graph')=\langle \phi(\Graph),\phi(\Graph') \rangle,
\end{equation}  
where $\phi(\Graph)$ denotes a vector of counts of substructures in $\Graph$ and $\langle \cdot , \cdot \rangle$ denotes an inner product in a reproducing kernel Hilbert space (RKHS). 

The two graph kernels to be introduced are both defined on labeled graphs. A graph is labeled if there is a mapping $l: \Vertices \to \Sigma$ where $\Sigma$ is a discrete set of labels.
\subsubsection{Shortest-path graph kernel}
The Shortest-Path kernel~\cite{borgwardt2005shortest} decomposes each graph $\Graph \in \mathcal{G}$ into shortest paths. Specifically, let $(l(v_a),l(v_b),s_{ab})$ denote a triplet where $s_{ab}$ signifies the length of the shortest path between two node $v_a,v_b$ and $l(v_a),l(v_b)$ are the labels of them, respectively. By collecting all the triplets in $\mathcal{G}$, one can get a list of unique triplets $T_{\mathcal{G}}$ and the Shortest-Path kernel can be defined as:
\begin{equation}
\mathcal{K}_{SP}(\Graph,\Graph')=\langle \phi_{T_{\mathcal{G}}}(\Graph),\phi_{T_{\mathcal{G}}}(\Graph') \rangle,
\end{equation} 
where $\phi_{T_{\mathcal{G}}}(\Graph)$ denotes a vector and the $i$-th element of it is the count of the $i$-th triplet $T^{(i)}_{\mathcal{G}}$ occurring in $\Graph$.
\subsubsection{Weisfeiler-Lehman graph kernel}
The Weisfeiler-Lehman kernel~\cite{shervashidze2009fast} decomposes each graph $\Graph \in \mathcal{G}$ into subtree patterns. Specifically, it replaces the label of each node in a graph with a compressed multiset label. The multiset label consists of the original label of the node and the sorted labels of its neighbors. This relabeling procedure is repeated for multiple times. By collecting all labels that occur at least once in one relabeling iteration of one graph of $\mathcal{G}$ as a list $L_{\mathcal{G}}$. The Weisfeiler-Lehman kernel can be defined as:
\begin{equation}
\mathcal{K}_{WL}(\Graph,\Graph')=\langle \phi_{L_{\mathcal{G}}}(\Graph) , \phi_{L_{\mathcal{G}}}(\Graph') \rangle,
\end{equation}
where $\phi_{L_{\mathcal{G}}}(\Graph)$ denotes a vector and the $i$-th element of it is the count of the $i$-th label $L^{(i)}_{\mathcal{G}}$ occurring in the relabeling iterations of $\Graph$.
\section{\ourwork}
\label{sec:ourwork}

As stated above, there are gaps between structural embedding and typical network embedding. We further summarize them as follows: 
	\begin{itemize}
	\item Existing network embedding methods treat nodes in anchor nodes' local neighborhoods as contexts. 
	Thus two anchor nodes residing far apart can hardly be related.
	\item To identify structural roles, one should focus on \textit{how} an anchor node connects with its context nodes, rather than \textit{which} nodes it links to. 
	However, typical network embedding methods focus on the latter. 
	\item Typical network embedding methods require some measure of smoothness (\eg, community memberships) over the network~\cite{donnat2018learning} while roles are not (to be shown in~\ref{sec:expressway}).
	\end{itemize}

To fill these gaps, existing structural embedding methods take a simple intuition: treating each node's local topology as a subgraph of the original graph, two nodes have similar roles if their local subgraphs are similar. One can come up with a naive solution based on this intuition: first use graph kernels to compute similarities between subgraphs, then construct a complete graph in which edge weights are proportional to subgraph similarities, finally employ network embedding methods on this weighted complete graph to learn node representations.

However, since typical graph kernels are built on graph level, subgraphs seen from perspectives of different nodes are not distinguished. \structovec corrected this by building a hierarchy based on a simplified rooted kernel.
Despite its success, \structovec still suffers time and space cost due to the pair-wise subgraph comparison and complete graph construction.
 
To tackle the above problems, we take the idea of relabeling and substructures from graph kernels and introduce a role identification procedure to better integrate the network embedding process.
Role identification traverses the subgraphs to capture the connecting patterns of the context nodes in their relation with the anchor node (\ie, roles of the context nodes in the subgraphs), and encode these patterns into new identifiers to get relabeled subgraphs. Thus a bridge from ``how'' to ``which'' is built.

In the relabeled subgraphs, context nodes which connect with their anchor nodes in the same manner will be treated as the same node even if they belong to different subgraphs. At the same time, structurally similar anchor nodes tend to have similar connecting patterns with their context nodes, and thus they share many context nodes in their relabeled subgraphs. Therefore, they are indirectly densely connected, and thus typical network embedding methods can be directly applied to learn structural embeddings.

Given a specific anchor node $v_i$ and a radius $k \leq k^*$, the role identification for a local subgraph $\Graph(\Nei_i^k)$ is a mapping function $\psi_i: \Nei_i^k \setminus \{v_i\} \to \mathcal{I}_i^k$, where $\mathcal{I}_i^k$ is a set of new identifiers. 

In the following subsections, we will propose two different role identification methods.

\subsection{\ourwork-SP}
\label{sec:ourworksp}
Firstly, we take similar intuition with \structovec and develop a role identification method named \ourwork-SP. 
We assume that two nodes are structurally similar if degrees of their neighbor nodes exhibit similar distribution. 
Similar to \structovec, which measures similarities for different neighborhood sizes, we take distances between context nodes and anchor nodes into account. 
Given a specific anchor node $v_i$ and a radius $k\leq k^*$, for a subgraph $\Graph(\Nei_i^k)$, we define the following role identification function:
\begin{equation}
\label{eq:ri-sp}
\psi_i(v_j)=h(\delta_i) \oplus h(\delta_j) \oplus s_{ij}, \;\; \text{for each} \; v_j \in \Nei_i^k \setminus \{v_i\},
\end{equation}
where $\oplus$ is the concatenation operator. 
To avoid the exponential growth in the number of new identifiers, we employ a discount function, which is defined as:
\begin{equation}
h(x)=\lfloor \log_2(x+1) \rfloor.
\end{equation}

Both $\delta_i$, $\delta_j$ and $s_{ij}$ are simple to measure, requiring no more than traversing (\eg, breadth-first searching) each member of $\Graph(\Nei_i^k)$. 

\subsubsection*{Relation with the Shortest-path kernel}
It can be observed that the new identifiers generated by \ourwork-SP are in line with the triplets in the Shortest-Path kernel. The difference between them is that \ourwork-SP is rooted and it can be viewed as having a labelling mapping defined as:
\begin{equation}
l(v_j)=h(\delta_j), \;\; \text{for each} \; v_j \in \Nei_i^k.
\end{equation}

\subsection{\ourwork-WL}
Since degree is an ambiguous measure of structural connectivities. 
\structovec and \ourwork-SP lack the ability to distinguish between context nodes that are at the same distance from $v_i$ and have the same degrees. 
Fig.~\ref{fig:role-identification} gives an illustrative example. As we can see, since node $b$ and $d$ have the same degree of 3 and are both at distance of 1 from the anchor node $a$, they can not be distinguished by \structovec and \ourwork-SP, even though $d$ is more densely connected with $a$ than $b$. 
Therefore, we develop the \ourwork-WL based on permutations of distances to capture fine-grained connectivity patterns.

Basically, there will always be multiple paths from the anchor node to each context node in its local subgraph, and permutations of these paths constitute the local topology. 
Thus, topological information (\eg, motifs) will be implied in the permutation of lengths of these paths.

For each context node, its relative position to the anchor node can be precisely revealed in the enumeration of all possible paths from the anchor node to it. 
However, enumerating all possible paths between two nodes is time consuming.
On the other hand, the simplest definition of relative position is the length of the shortest path but the definition is too broad. 
We trade off between the above two definitions and develop a neighborhood aggregation process to get the relative positions of the context nodes.

Given a specific anchor node $v_i$ and a radius $k\leq k^*$, for a subgraph $\Graph(\Nei_i^k)$, we let each context node $v_j \in \Nei_i^k$ correspond to a vector $\nx_{ij}$ of length $k+1$. 
Using $\nx^{(n)}$ to denote the $n$-th element of $\nx$, we set
\begin{equation}
\nx_{ij}^{(n)} = \left| \{v_l \in \Nei_j \mid s_{il}=n\} \right|, \;\; \textrm{for} \; n\in \{0,1,\ldots,k\}.
\end{equation}
In other words, the $n$-th element of $\nx_{ij}$ denotes the number of neighbor nodes of $v_j$ whose distance from $v_i$ is exactly $n$.

Then we define a role identification function as follows:
\begin{equation}
\label{eq:ri-wl}
\begin{split}
\psi_i(v_j)=h(\nx_{ii}) \oplus h(\nx_{ij}) \oplus s_{ij}, \;\; \text{for each}\; v_j \in \Nei_i^k \setminus \{v_i\},
\end{split}
\end{equation}
where $h$ is the same discount function as in \ref{sec:ourworksp}.

\subsubsection*{Relation with the Weisfeiler-Lehman kernel}
\ourwork-WL takes the idea of \textit{neighborhood aggregation} from the Weisfeiler-Lehman kernel. The differences between them are 
1) \ourwork-WL is rooted, 
2) \ourwork-WL takes a labelling mapping defined as:
\begin{equation}
l(v_j)=s_{ij}, \;\; \text{for each} \; v_j \in \Nei_i^k,
\end{equation}
3) \ourwork-WL counts the labels of the neighbors instead of sorting them, 
4) \ourwork-WL takes discount on the counts when compressing the multiset label, 
5) \ourwork-WL takes only one iteration.

\subsection{Random-walk-based Embedding}

Given a role identification method defined as above, we are able to get a relabeled subgraph $\tilde \Graph(\Nei_i^{k})$ for each anchor node $v_i$.
In this newly generated graph, only the anchor node $v_i$ keeps its original identifier, while the other context nodes get new ones.

For parallelizability and simplicity, we adopt random-walk-based network embedding approaches to learn representations.
Specifically, for each subgraph $\tilde \Graph(\Nei_i^{k})$, we perform fixed-length random walks starting from the anchor node $v_i$ for a certain number of times. 
Note that random walk on each subgraph can be performed in parallel.

Through random walk simulating, the structural dependencies between substructures (\ie, new identifiers) are kept in random walks. 
By merging all random walks together as a corpus, we can train a language model to learn node embeddings by treating node sequences as sentences. 
For this work, we use the Skip-Gram model with negative sampling~\cite{mikolov2013efficient,mikolov2013distributed}.

\subsection{The \ourwork\ Algorithm}
\begin{algorithm}[t]
\textbf{Input:} Graph $\Graph = (\Vertices,\Edges)$, Neighborhood size $k$, Dimensions $d$, Walks per node $\gamma$, Walk length $\lambda$, Window size $\omega$
\begin{algorithmic}[1]
\STATE Initialize walks $W$ to $\emptyset$ \label{algline:init}
\FOR{$i = 1,2,...,\lvert \Vertices \rvert$}    \COMMENT{\textbf{parallel walks per node}}
\label{algline:walkpernode} 
	\STATE $\tilde \Graph(\Nei_i^k)=$ {RoleIdentificationSubgraph}($\Graph(\Nei_i^k)$)
	\STATE $W_i=$ {RandomWalkOnSubgraph}($\tilde \Graph(\Nei_i^k)$, $\gamma$, $\lambda$)
	\STATE Add $W_i$ to $W$
\ENDFOR \label{algline:main_endfor}
\STATE $f = \text{SkipGram}$($W$, $d$, $\omega$) \COMMENT{\textbf{parallel}}
\label{algline:sgd} 
\STATE \textbf{return} the learned node embeddings $f$ \label{algline:return} 
\end{algorithmic}
\caption{\ourwork($\Graph$, $k$, $d$, $\gamma$, $\lambda$, $\omega$)}
\label{alg:ourwork}
\end{algorithm}

The pseudocode of \ourwork is present in Algorithm~\ref{alg:ourwork}. 
\ourwork takes a graph as input, and outputs low-dimensional representation for each node in the graph. 
Nodes with similar roles will have similar representations. 
For each node, \ourwork generates a relabeled subgraph for it and simulates random walks on the generated subgraph. 
By putting all random walks together and learning a language model, the embeddings can be obtained.
It is worth noting that both the role identification procedure (line~\ref{algline:walkpernode}) and the network embedding procedure (line~\ref{algline:sgd}) are parallelizable, making \ourwork suitable for large networks.

\subsection{Complexity Analysis}
\subsubsection{Time complexity}
Given a subgraph $\Graph(\Nei_i^k)$ rooted at an anchor node $v_i$, computation of $s_{ij}$ for each context node $v_j$ needs to traverse each edge in the subgraph. 
Suppose $k=k^*$, which means there is $\Graph(\Nei_i^k) \equiv \Graph$ for each anchor node $v_i$. Thus traversing all subgraphs takes a complexity of $O(\lvert \Vertices \rvert \cdot \lvert \Edges \rvert)=O(\bar{\delta}\lvert \Vertices \rvert^2)$, where $\bar{\delta}$ is the average degree of the nodes in the graph. 
Then the final complexity of \ourwork-SP in the worst case is $O(\bar{\delta} \lvert \Vertices \rvert ^2)$. 
Since \ourwork-WL needs to traverse each neighbor of the context node to get the vector $\nx$, the complexity of \ourwork-WL in the worst case is $O(\bar{\delta}^2 \lvert \Vertices \rvert ^2)$ 
\subsubsection{Space complexity}
\ourwork-SP needs to store the mapping $\phi$ and \ourwork-WL needs to store the vectors $\nx$, both of which require space proportional to $\Nei_i^k$.
Suppose $k=k^*$, which means there is $\Graph(\Nei_i^k) \equiv \Graph$ for each anchor node $v_i$.
Since the role identification procedure is performed parallelizably on subgraphs, the final space complexity of \ourwork-SP and \ourwork-WL in the worst case is $O(\lvert \Vertices \rvert )$

\section{Experiments}
\label{sec:experiments}
We conduct extensive experiments to prove the effectiveness and efficiency of \ourwork.
The experiments are carried out on a single machine with 32GB RAM, 16 CPU cores at 3.4GHz using 10 threads (without GPU support).  
The source code of \ourwork is available online\footnote{https://github.com/maxuewei2/RiWalk}.

\subsection{Compared Algorithms}
We compare our algorithm with the following state-of-the-art embedding methods.
\begin{itemize}
\item \nodetovec : A typical network embedding method which adopts biased random walks and language models to learn node representations. The biased random walk introduces flexibility in neighborhood exploring to learn richer embeddings. Default parameter settings for this approach are in line with the typical values decribed in the paper, \ie, dimensions of embedding $d=128$, number of walks per node $\gamma=10$, walk length $\lambda=80$, window size $\omega=10$. The hyper-parameter $p,q$ are fine-tuned in all our following experiments.

\item \structovec: A structural embedding approach which learns structural node representations based on degree distributions in the neighborhoods. Default parameter settings for this approach are in line with \nodetovec for fair comparison, \ie, dimensions of embedding $d=128$, number of walks per node $\gamma=10$, walk length $\lambda=80$, window size $\omega=10$. 
It is worth noting that all the three optimizations of \structovec described in the paper are used in our following experiments. 
As a result, the number of layers of the hierarchy $k$ in \structovec becomes a hyper-parameter to be tuned.
We set $k=4$ as default.

\item \graphwave: A structural embedding method which treats diffusion kernels as probability distributions over the graph and characterzes the distributions to obtain structural embeddings. 
By default, we set dimensions of embedding $d=128$ and use the multiscale version with evenly spaced sampling points in range [0, 100]. 
\item Majority: This naive method simply predicts the labels that occur most in training data as output.
\end{itemize}

To make fair comparisons, the default parameters for our methods are set to: number of walks per node $\gamma=80$, walk length $\lambda=10$, window size $\omega=10$, neighborhood size $k=4$. Thus \nodetovec, \structovec and \ourwork generate equal number of samples in the sampling phase.

It is worth noting that we exclude \rolx~\cite{henderson2012rolx} since it has been shown to be inferior to \structovec~\cite{ribeiro2017struc2vec} and \graphwave~\cite{donnat2018learning}.

\subsection{Case Study: Expressway Network}
\label{sec:expressway}

As stated in Section~\ref{sec:ourwork}, typical network embedding methods require some measure of smoothness  over the network. Here we  empirically demonstrate the fact by giving an illustrative example. 
\subsubsection{Data and setup}

\begin{figure}[t]
\centering    
\subfigure[Expressway network] {
\label{fig:expressway-graph}     
\includegraphics[width=0.45\columnwidth]{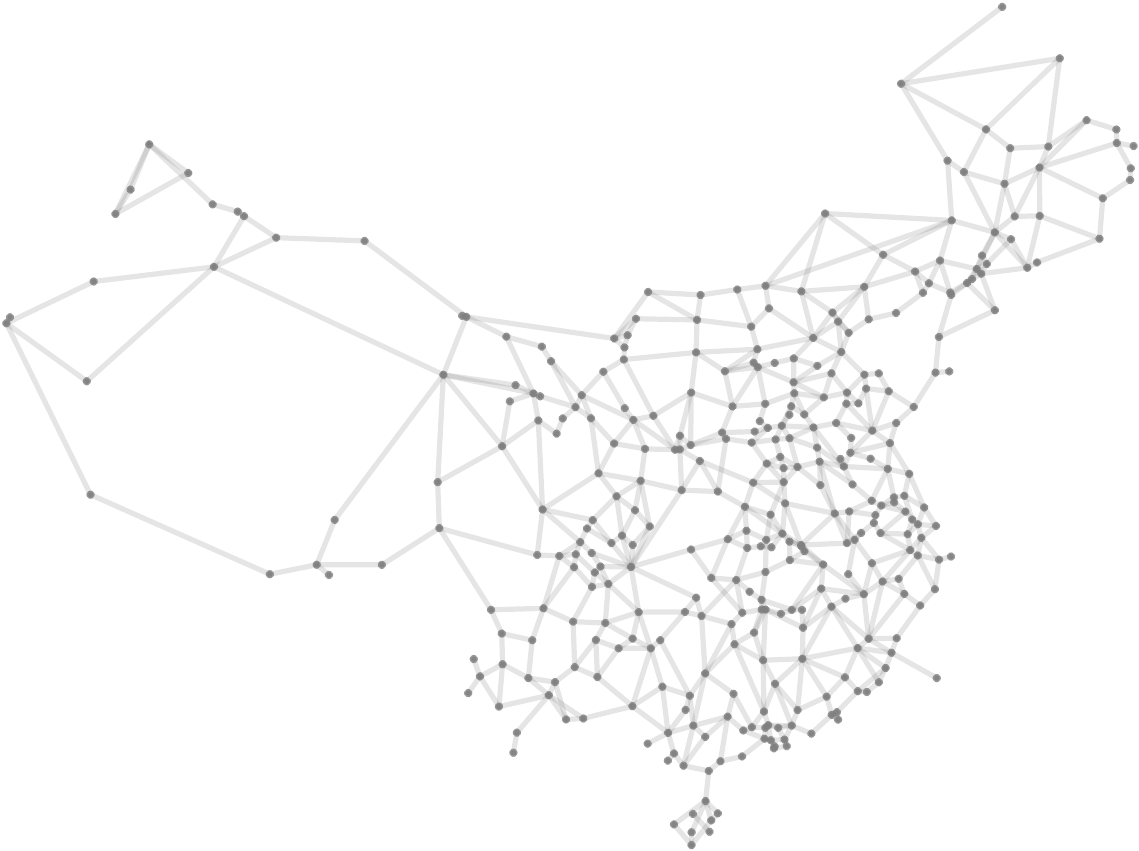}
}
\subfigure[Distribution of labels corresponding to geographical regions of cities] {
\label{fig:expressway-gr}
\includegraphics[width=0.45\columnwidth]{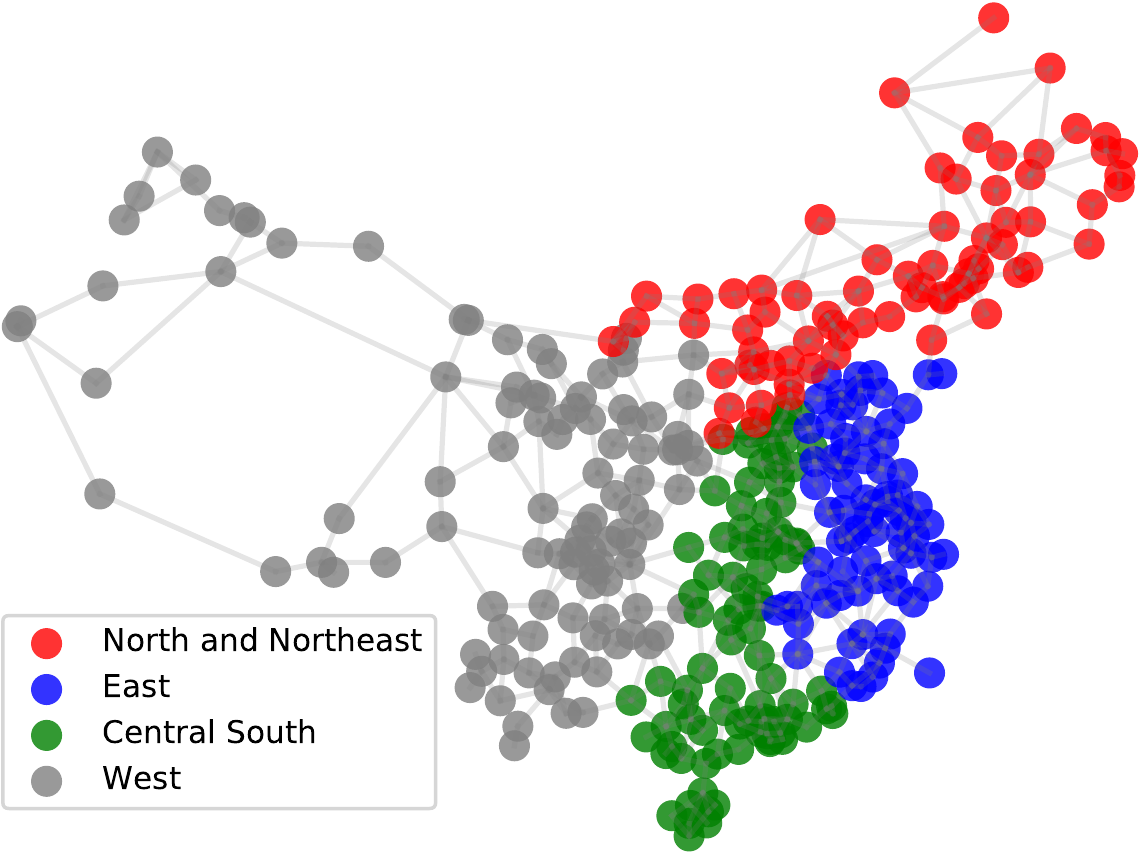}
}
\\
\subfigure[Distribution of labels corresponding to statuses of cities] {
\label{fig:expressway-status}
\includegraphics[width=0.45\columnwidth]{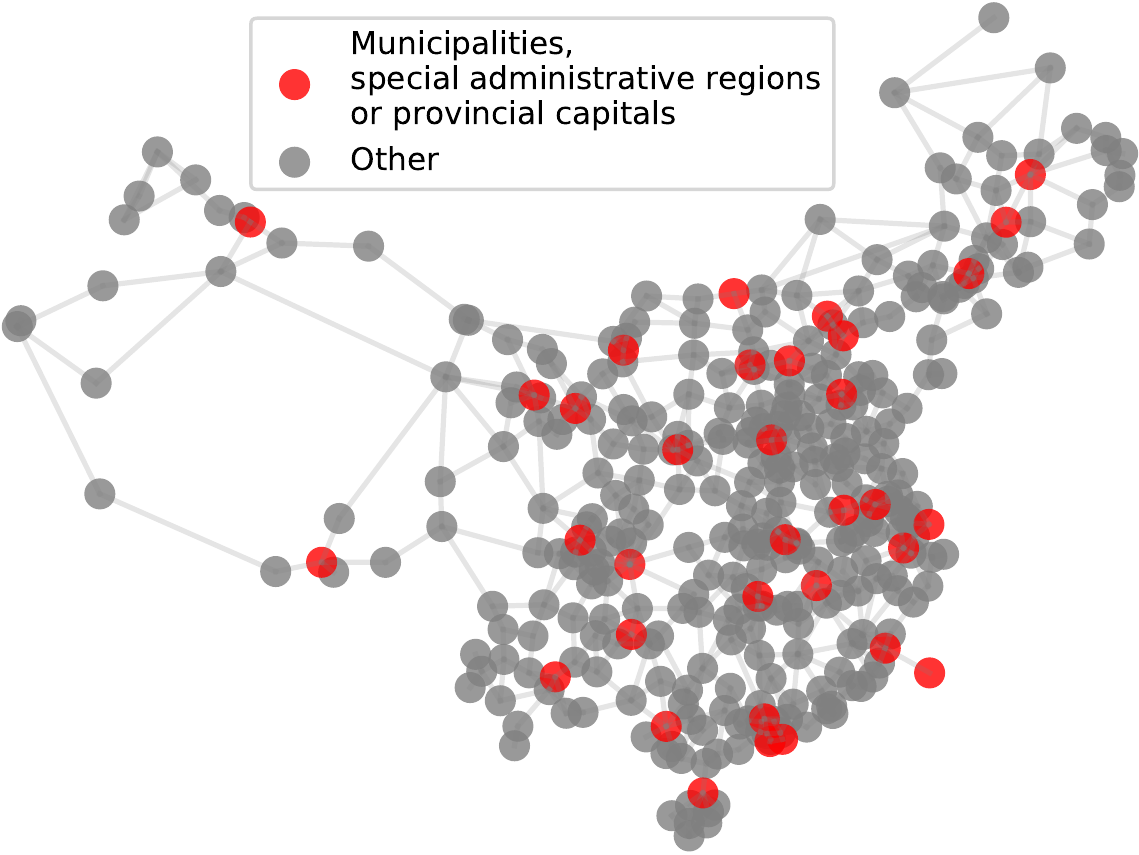}
}
\subfigure[Distribution of labels corresponding to number of stores of Starbucks of cities] {
\label{fig:expressway-starbucks}
\includegraphics[width=0.45\columnwidth]{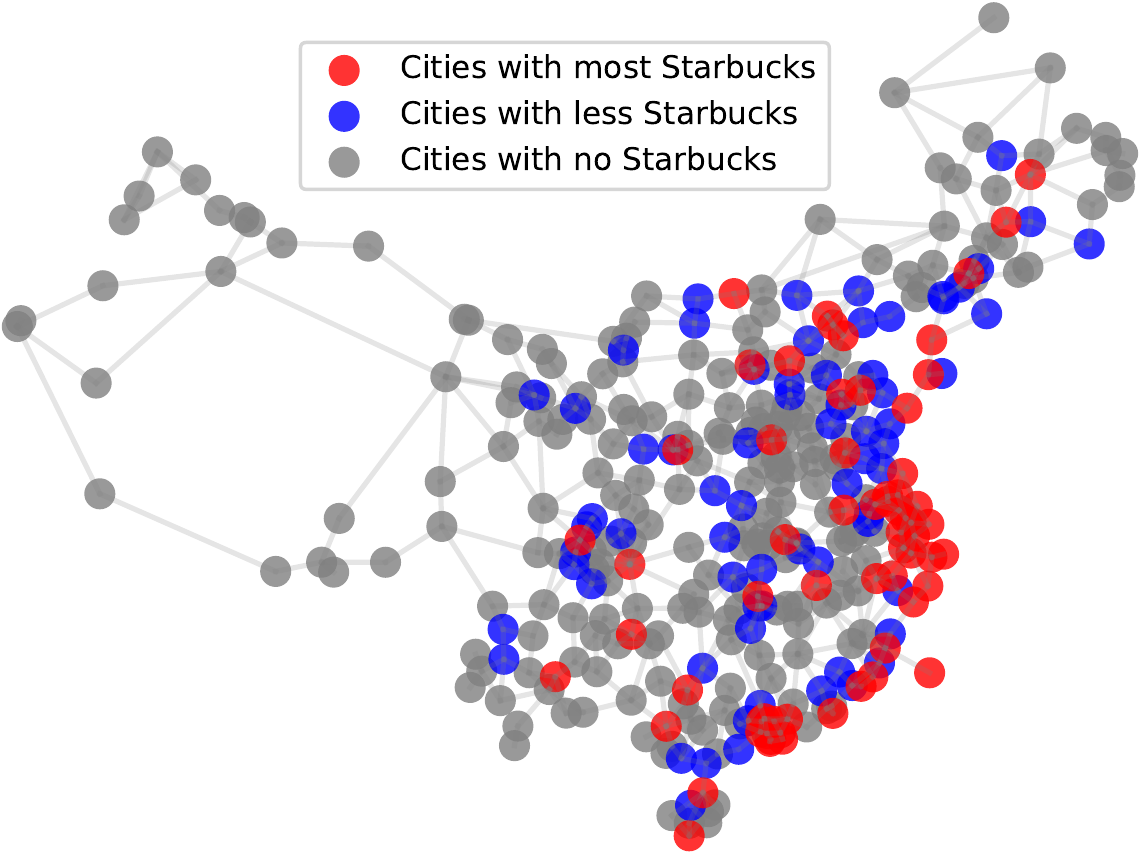}
}
\caption{Label distributions over the Expressway network.}
\label{fig:expressway}
\end{figure}

We collect data from Wikipedia pages under the category ``Chinese national-level expressways''\footnote{https://wikipedia.org/wiki/Category:Chinese\_national-level\_expressways} to construct an undirected, unweighted expressway network (shown in Fig.~\ref{fig:expressway-graph}). 
Nodes in the network are cities and edges indicate the existence of expressways between the cities. The network has 348 nodes and 675 edges. It is worth noting that the diameter of this network is 24.

We assign class labels to the nodes in three different ways to get three datasets:
\begin{itemize}
\item Expressway geo. region (Fig.~\ref{fig:expressway-gr}): The cities are split into four classes based on their geographical regions. 
Thus the class labels indicate community memberships.
\item Expressway status (Fig.~\ref{fig:expressway-status}): The cities are split into two classes based on their statuses:
	\begin{enumerate*}
	\item municipalities, special administrative regions or provincial capitals, 
	\item neither of the former three.
	\end{enumerate*}
	Thus the class labels are related to the roles (\ie, administrative levels) of the cities.
\item Expressway Starbucks (Fig.~\ref{fig:expressway-starbucks}): Class labels are assigned based on the number of stores of Starbucks in the cities. Specifically, for the cities which have Starbucks stores, we use the median number to divide them into two groups: cities with more Starbucks and cities with less Starbucks. Then the third group is formed by the cities with no Starbucks.
\end{itemize}

Since this is a road network, it is flat and we can visually see the distributions of the class labels over the network. Imagine that different colors denoting different hights. As we can see, labels corresponding to geographical regions are smoothly distributed over the network while labels corresponding to statuses are completely not. 
Besides, labels based on numbers of Starbucks stores are semi-smoothly distributed. For example, cities with most Starbucks (the red ones in Fig.~\ref{fig:expressway-starbucks}) are either southeastern coastal cities or capitals of provinces, and the coastal cities are distributed smoothly while the provincial capitals are not.

For each of the three datasets, we perform a multi-class node classification task on the network \wrt the correponding assigned labels. Specifically, for each dataset, we learn latent representations of nodes of the network using different embedding approaches. Then for each approach, a portion ($T_R$=80\%) of node embeddings along with their labels are randomly chosen to be training data. The rest node embeddings and their labels are used as test. We feed the training data to a one-vs-rest logistic regression classifier with L2 regularization (implemented by LibLinear~\cite{fan2008liblinear}). 
Then the trained classifier is used to predict the labels of the test feature embeddings.
We take the $\text{F}_1$ score as the performence metric since the classes of the three datasets are both imbalanced. Note that each individual embedding is evaluated 10 times to reduce the noise introduced by classifier.
This embedding process is repeated for 10 times and the average performance is reported.
The parameter settings for each approach on each dataset are tuned to be optimal by using grid search.
We report the $\text{Macro-F}_1$ scores in Table~\ref{tab:expressway}. The results of $\text{Micro-F}_1$ scores are omitted since they follow a similar trend.

\begin{table}[t]
\caption{Results of classification tasks on Expressway networks. ($\text{Macro-F}_1$ (\%) )}  
\label{tab:expressway}
\centering
\begin{tabular}{lccc}
\toprule
\multirow{3}{*}{\textbf{Algorithm}} & \multicolumn{3}{c}{\textbf{Dataset}}\\ 
                                      \cmidrule(lr){2-4}
             & Expressway & Expressway & Expressway \\
             & geo. region & status & Starbucks \\ 
             \midrule
\nodetovec &\textbf{96.13} & 50.09 & \textbf{52.75}\\
\structovec &39.78 & \textbf{54.44} & 38.60\\
\graphwave &\textbf{51.39} & 51.28 & 42.88\\
\ourwork-SP &50.05 & 53.05 & 44.26\\
\ourwork-WL &\textbf{51.39} & \textbf{56.62} & \textbf{44.70}\\
\midrule
Majority &11.41 & 47.37 & 25.37\\
\bottomrule
\end{tabular}
\end{table}

\subsubsection{Results}
As we can see, the classification accuracy is highly related to the smoothness of the label distribution. 
In particular, when class labels distribute smoothly over the network (\ie, the Expressway geo. region dataset), the typical network embedding method (\ie, \nodetovec) gives the best performance. When such a kind of smoothness does not exist (\ie, the Expressway status dataset), typical network embedding method performs the worst. And in this case, since the class labels are related to the roles of the nodes, all the structural embedding methods perform better than \nodetovec. In the third case, \nodetovec and \ourwork-SP achieve 107.9\% and 76.19\% gain over the Majority method, respectively. We argue that this is a consequence of the semi-smooth distribution of the class labels over the network. \nodetovec leverages the smooth part of it and structural embedding methods leverage the non-smooth part. This case also shows potencial of combining both typical network embedding and structural embedding, where the former one captures community memberships and the latter one captures structural roles.

\subsection{Within-Network Role Classification}
\label{subsec:classification}

We conduct the node classification task on the following datasets.
\begin{itemize}
\item European air-traffic network (shortly Europe) and American air-traffic network (shortly USA)~\cite{ribeiro2017struc2vec}: Both the two networks are air-traffic networks, with nodes indicating airports and edges representing commercial flights between airports. 
Class labels indicate the levels of activity 
(\eg, the total number of landings plus takeoffs in a corresponding period)
 of the airports, thus the labels are related to the roles of the nodes.

\item English-language film network (shortly Film) ~\cite{tang2009social}: This is a film-director-actor-writer network, with each node assigned a label indicating it belonging to one of the four roles.  
The edges between nodes denote co-occurances on the same Wikipedia page. 

\item Actor co-occurance network (shortly Actor) ~\cite{tang2009social}: This is the actor only subgraph of the Film network. 
We sort the nodes based on the number of words of their Wikipedia pages, and then use quartiles to split the nodes into four groups.
Thus the group labels can be seen as a measure of the influences of the nodes.
\end{itemize} 
All these networks are undirected and unweightecd. The detailed statistics of these datasets are summarized in Table~\ref{tab:datasets}.

\begin{table}[h]
\caption{Detailed statistics of datasets used for classification.}  
\label{tab:datasets}
\centering
\begin{tabular}{lcccc}%@{\hskip 2mm} @{}l@{\extracolsep{-2mm}}ccccccc@{}
\toprule
                             & Europe & USA   & Film   & Actor \\ \midrule
\# Vertices     & 399    & 1190  & 27312  & 7779  \\
\# Edges      &5995    & 13599 & 122514 & 26752 \\
\# Classes  & 4      & 4     & 4      & 4 \\
\bottomrule
\end{tabular}
\end{table}

As the name suggests, in the within-network classification task setting, the training and the testing data are from the same network.
In fact, all the node classification task carried out in \ref{sec:expressway} are within-network classification tasks.
Here we follow the same experimental procedure as in \ref{sec:expressway} except that we range the training ratio ($T_R$) from 10\% to 90\%.
The parameter settings used for the USA network are tuned to be optimal by using grid search.
And the parameters used for the Film and the Actor network are default settings. 
The average Micro-F$_1$ scores are reported.

\begin{table*}[t]
\caption{Micro-F$_1$(\%) scores of within-network role classification.}  
\label{tab:cls}
\centering
\begin{tabular}{@{\extracolsep{\fill}}llccccccccc |lll}%@{\hskip 2mm}l@{\hskip 3mm} 
\toprule
        &          & \multicolumn{9}{c}{Labeled Nodes (\%)}      & \multicolumn{3}{c}{Time and Memory Usage}   \\ 
                     \cmidrule(lr){3-11}                           \cmidrule(lr){12-14} 
Dataset &  Method  & 10 & 20 & 30 & 40 & 50 & 60 & 70 & 80 & 90  & Mem (M) &Real (s) &User (s)  \\ 
\midrule
\multirow{5}{*}{USA} 
 & \nodetovec & 54.86 & 58.84 & 61.03 & 61.78 & 62.79 & 63.44 & 63.74 & 63.86 & 64.18 &  &  & \\
 & \structovec & 54.39 & 58.06 & 60.23 & 60.93 & 61.86 & 62.73 & 63.17 & 64.38 & 65.75 & 82 & 94 & 863\\
 & \graphwave & \textbf{60.30} & \textbf{61.30} & \textbf{62.45} & 62.90 & 62.38 & 62.98 & 62.36 & 63.25 & 64.67 & 127 & 6 & 74\\
 & \ourwork-SP & 58.62 & 60.35 & 61.21 & 63.03 & 63.69 & 63.58 & 64.47 & 65.83 & 64.60 & 13 & 4 & 19\\
 & \ourwork-WL & 58.25 & 60.82 & 62.39 & \textbf{63.04} & \textbf{64.34} & \textbf{64.38} & \textbf{65.92} & \textbf{66.17} & \textbf{66.25} & 42 & 17 & 146\\
\midrule
\multirow{5}{*}{Film} 
 & \nodetovec & 44.04 & 45.36 & 45.91 & 46.10 & 46.36 & 46.33 & 46.46 & 46.75 & 46.68 &  &  & \\
 & \structovec & 54.14 & 55.59 & 56.10 & 56.24 & 56.37 & 56.54 & 56.46 & 56.70 & 56.43 & 1027 & 1972 & 18236\\
 & \graphwave & — & — & — & — & — & — & — & — & — & — & — & —\\
 & \ourwork-SP & \textbf{60.26} & \textbf{61.08} & \textbf{61.40} & \textbf{61.52} & \textbf{61.61} & \textbf{61.63} & \textbf{61.65} & \textbf{61.44} & \textbf{61.56} & 111 & 179 & 1148\\
 & \ourwork-WL & 59.15 & 60.23 & 60.48 & 60.71 & 60.67 & 60.82 & 60.76 & 60.88 & 60.98 & 113 & 600 & 5404\\
\midrule
\multirow{5}{*}{Actor} 
 & \nodetovec & 31.24 & 33.34 & 34.88 & 35.74 & 36.04 & 36.83 & 36.61 & 37.14 & 37.82 &  &  & \\
 & \structovec & 42.46 & \textbf{44.72} & \textbf{45.43} & \textbf{45.99} & \textbf{46.51} & \textbf{46.56} & \textbf{47.05} & \textbf{47.48} & \textbf{47.56} & 284 & 379 & 3459\\
 & \graphwave & — & — & — & — & — & — & — & — & — & — & — & —\\
 & \ourwork-SP & \textbf{43.27} & 44.61 & 45.05 & 45.60 & 45.31 & 45.78 & 46.56 & 46.05 & 45.13 & 65 & 30 & 177\\
 & \ourwork-WL & 41.60 & 43.43 & 44.25 & 44.16 & 44.69 & 45.27 & 45.39 & 45.23 & 46.54 & 64 & 62 & 545\\

\bottomrule
\end{tabular}
\end{table*}

We summarize the results of within-network node classification in Table~\ref{tab:cls}. Highest performances are shown in bold.
As we can see, \ourwork-SP and \ourwork-WL outperform the other approaches on the USA dataset when the training ratio $T_R \geq 40\%$. On the Film and the Actor network, \graphwave runs out of memory when running the algorithm and the other three structural embedding methods outperform \nodetovec. This indicates that typical network embedding methods are not capable of identifying structural roles of nodes. When comparing between structural embedding methods, on the Film network, \ourwork-SP and \ourwork-WL perform the best and give us 9\% and 8\% gain over \structovec, respectively. On the Actor network, \ourwork-SP and \ourwork-WL achieve comparable performance with \structovec.

Besides, we also notice that \structovec performs inferior to all the other structural embedding methods when $T_R$ is small. For example, on the USA dataset, even though \structovec achieves comparable performance with the other approaches when $T_R \geq 70\%$, it performs worst when $T_R=10\%$. And on the Actor network, even if \structovec gives the best performance when $T_R \geq 20\%$, it perfroms worse than \ourwork-SP when $T_R\leq 10\%$.
Since \ourwork-SP shares similar intuition with \structovec, this phenomenon demonstrates the effectiveness of our framework, which provides robust performance on sparsely labeled graphs.

\subsubsection{Time and space usage analysis}
We record the running time and the memory usage of each approach in the within-network classification tasks.
Since all algorithms were implemented using Python, we are able to make more fair comparisons on algorithm level. The results are listed on the right side of Table~\ref{tab:cls} (\textit{Real} refers to actual elapsed time; \textit{User} refers to CPU time used only by the process).
Since \nodetovec is a typical network embedding algorithm, it is ommited when comparing.

As shown in the table, \graphwave uses 10x more memory than \ourwork-SP and this memory usage prevents it from running on the two larger networks. Besides, on the Film network, \structovec also uses 10x more memory than \ourwork-SP and \ourwork-WL.

On the Film network, \ourwork-SP and \ourwork-WL use 11x and 3x less time than \structovec \wrt real time usage, respectively. On the Actor network, \ourwork-SP and \ourwork-WL use 20x and 6x less time than \structovec \wrt user time usage, respectively.

The above results demonstrate the effciency of our proposed framework, which achieves comparable performances while being an order of magnitude more efficient in terms of both time and space.

\begin{figure}[t]
\centering    
\subfigure[USA] {
\label{fig:pa-usa}     
\includegraphics[width=0.45\columnwidth]{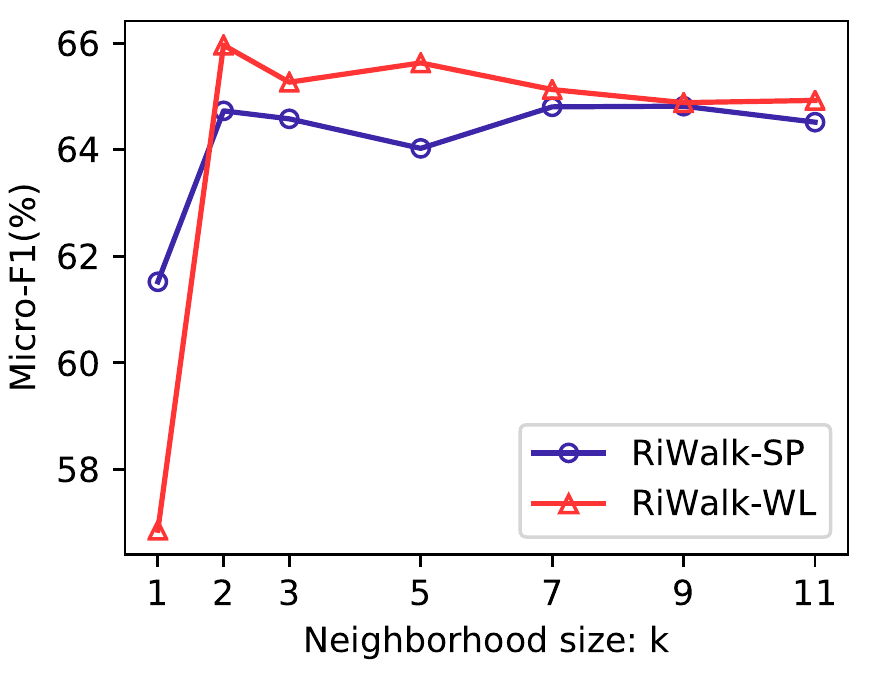}
}
\subfigure[Actor] {
\label{fig:pa-actor}
\includegraphics[width=0.45\columnwidth]{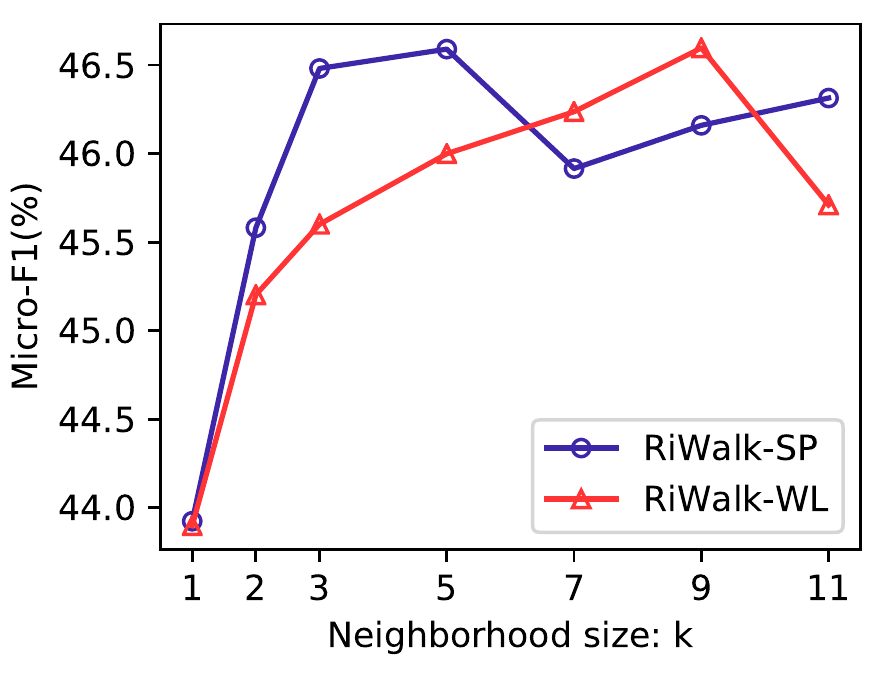}
}
\caption{Performance \wrt neighboorhood size $k$.}
\label{fig:pa}
\end{figure}

\subsubsection{Parameter study}
We study the performance \wrt the neighborhood size $k$. Specifically, we set $T_R=80\%$ and range $k$ in different values while keep the other parameters unchanged. The results are reported in Fig.~\ref{fig:pa}.
We can see that the performances saturate or even get worse when $k$ keeps increasing. We argue that this is the consequence of the noise introduced by larger neighborhood. Besides, we also notice that the algorithm can achieve a satisfying performance when $k$ is fairly small, \eg, $k=2$ on the USA dataset, or $k=3$ on the Actor dataset. This fact further reduces the complexity of the algorithm when applying on real-world networks.
\subsection{Across-Network Role Classification}

\label{sec:transfer}
To test the potential of \ourwork in transfer lerning, we study the problem of role classification across networks. 
In the across-network classification setting, one network along with all the class labels of its nodes are treated as training data, and the task is to predict the class labels of the nodes in another network. Obviously, the key problem of this task is to capture the features that transfer across networks.

Here we consider the problem of identifying hub nodes in different networks. 
That is, given a network whose nodes are already classified as hub nodes or non-hub nodes, we aim to leverage this knowledge to determine if one node in another network is a hub.

We use the Europe, USA and Actor network to create four merged-network datasets: First, for each network, we treat the nodes of the first group (\eg, hub airports, famous actors) as hub nodes and the other three groups as non-hub nodes. We treat these binary class labels as ground-truth facts. Then, given two networks $\Graph_a$ and $\Graph_b$, we construct a network $\Graph_{a:b}$ composed of the two networks. The merged network then is fed to different embedding approaches to learn node embeddings. Treating embeddings of nodes in $\Graph_a$ along with their class labels as training data, we train a logistic regression classifier to predict the labels of all nodes in $G_b$. The parameter settings for different approaches are tuned to be optimal. 
The experiments are repeated for 10 times and we report the average Macro-F$_1$ score among the runs.

\begin{table}[t]
\caption{Macro-F$_1$(\%)) scores of across-network role classification.}  
\label{tab:trans}
\centering
\begin{tabular}{lcccc}
\toprule
\multirow{2}{*}{\textbf{Algorithm}} & \multicolumn{4}{c}{\textbf{Dataset}} \\
                                      \cmidrule(lr){2-5}
                                 & USA:Europe &Europe:USA &Actor:USA &USA:Actor    \\ \midrule
\nodetovec &42.92 & 45.99 & 46.91 & 42.88\\
\structovec &78.87 & \textbf{79.74} & \textbf{80.13} & 57.48\\
\graphwave &\textbf{86.17} & 73.98 & — & —\\
\ourwork-SP &\textbf{81.98} & \textbf{80.07} & 78.95 & \textbf{73.97}\\
\ourwork-WL &81.95 & 78.99 & \textbf{80.90} & \textbf{67.34}\\
\midrule
Majority &42.91 & 42.87 & 42.87 &42.86\\
\bottomrule
\end{tabular}
\end{table}

As shown in Table~\ref{tab:trans}, \nodetovec gives close performance with the Majority method. This indicates that typical network embeddings are not able to transfer across networks.
At the same time, all the structural embedding methods show much better performances, which demonstrates the potential of structural embeddings in transfer learning. 

As we can see, \ourwork-SP and \ourwork-WL give the best performance among the four datasets.
Specifically, on the USA:Europe and the Actor:USA datasets, \ourwork-SP and \ourwork-WL show competitive performances with our baselines. And on the Europe:USA dataset, \ourwork-SP and \ourwork-WL give us 8.2\% and 6.8\% gain over \graphwave, respectively. On the USA:Actor dataset, \ourwork-SP and \ourwork-WL give us 28.7\% and 17.2\% gain over \structovec, respectively.

Besides, we notice that the performances of nearly all the structural embedding methods drop significantly on the Europe:USA and the USA:Actor datasets compared to the other two datasets.
We argue that this is the consequence of the lack of knowledge when transferring from small networks to large networks. However, despite this lack of knowledge, \ourwork-SP and \ourwork-WL give much better performances than our baselines. This again shows the robustness of our proposed framework.

It is also worth noting that the three networks used in this experiment (Europe, USA and Actor) have different sizes and they are from different scenarios: infrastructure and social network, respectively.
Thus this experiment also shows the universality of structural roles among different network regardless of their sizes and scenarios.

\subsection{Structural Hole Identifiaction}

\begin{figure}[t]
  \centering
  \includegraphics[width=0.8\linewidth]{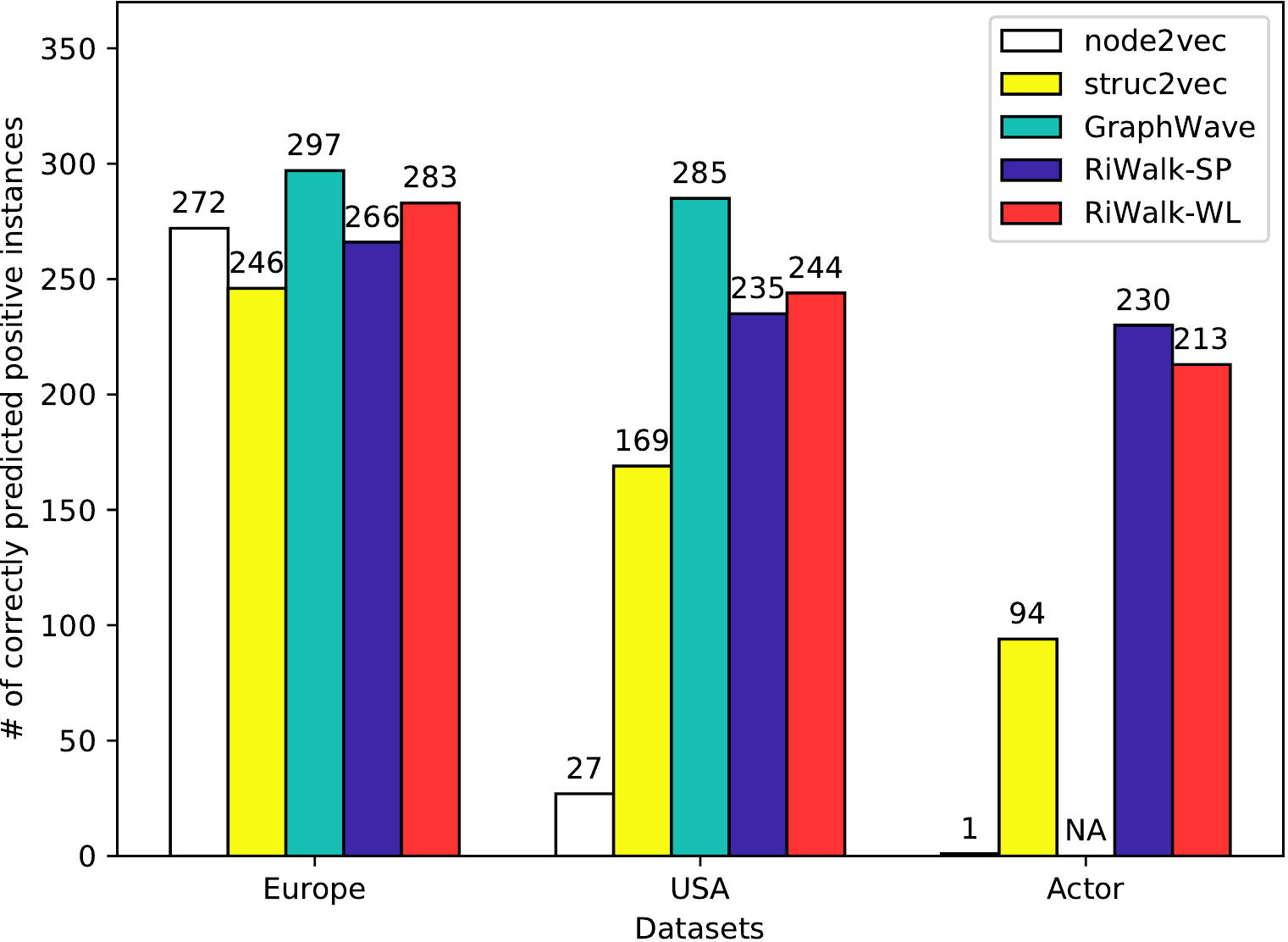}
  \caption{Results of structural hole identification.}
\label{fig:st}
\end{figure}

Structural holes~\cite{burt2009structural} play an important role in the information exchange process over the network.
Identifying structural holes in networks would help researcher to gain a thourough understanding of network evolution and information cascades.

Here we take the Europe, USA and Actor network as our datasets.
Given a network, we first compute the constraint on all nodes in the network.
We take the constraint values as ground-truth measurements of nodes bridging structural holes.
That is, the lower constraint value one node has, the more opportunities it has for bridging structural holes.
Then the first 300 nodes with smallest constraint values are chosen to be positive instances.
We then learn node embeddings on the network with different approaches. 
Denoting the node with the smallest constraint value as $v_{\text{sh}}$, for each approach, we compute cosine similarities between $v_{\text{sh}}$ and all nodes in the network using their embeddings.
The first 300 nodes that have most similar embeddings with $v_{\text{sh}}$ are treated as predicted positive instances. The number of correctly predicted instances are reported in Fig.~\ref{fig:st}. The parameter settings for each approach are tuned to be optimal.

As we can see, thanks to the flexibility of neighborhood exploring introduced by biased random walk, \nodetovec can capture structural equivalence in small networks (\ie, the Europe network). 
However, when it comes to larger networks, the performance of \nodetovec drops significantly.
Similar performance drops are also observed on the four structural embedding methods. 
\structovec drops the most, and as a result, \ourwork-SP and \ourwork-WL give us 145\% and 127\% performance gains over \structovec on the Actor network, respectively. This experiment once again shows the robustness of our proposed algorithms.

\subsection{Scalability}
\begin{figure}[t]
  \centering
  \includegraphics[width=0.8\linewidth]{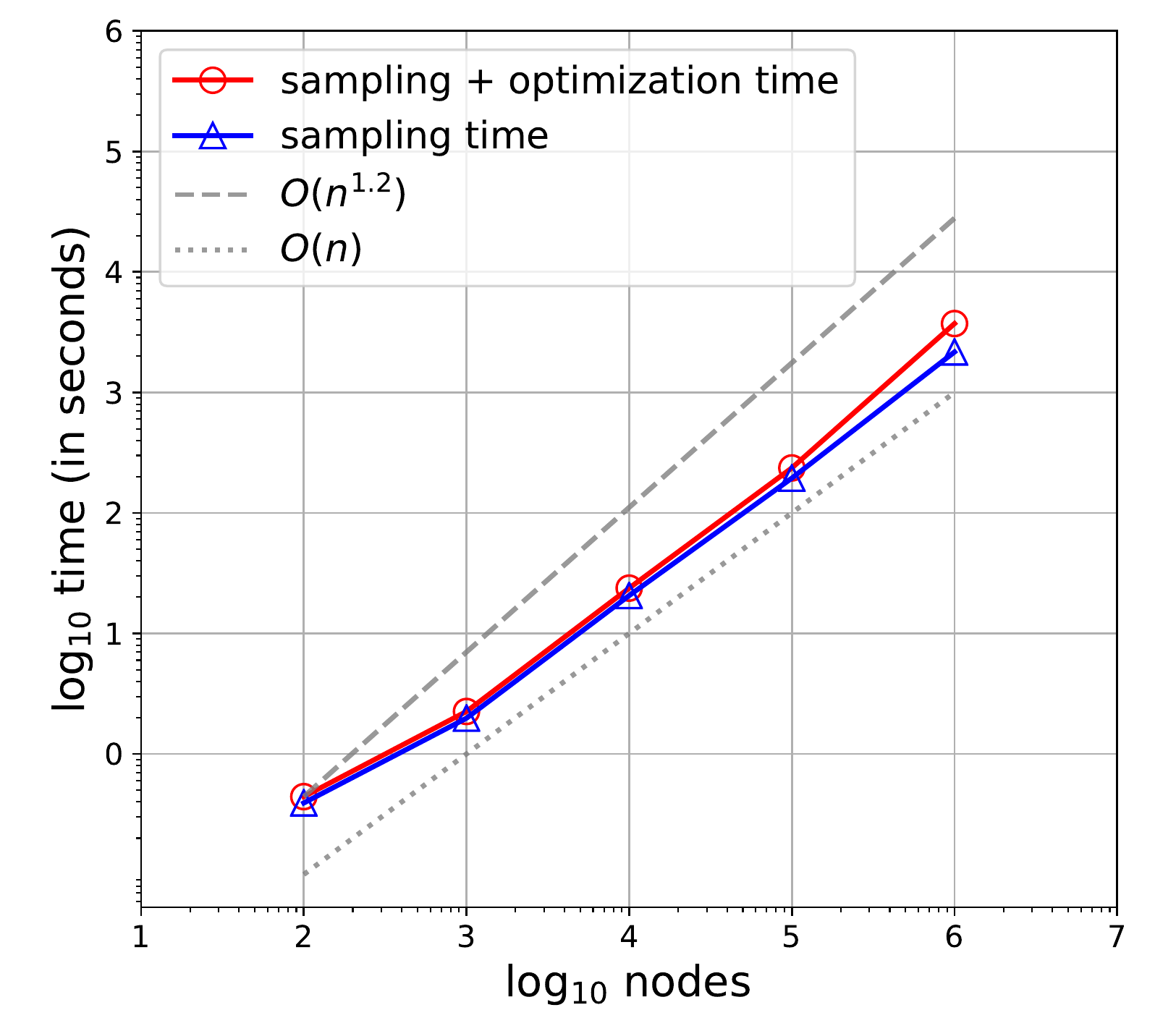}
  \caption{Running time on Erdos-Renyi graphs.}
\label{fig:scale}
\end{figure}

To evaluate the scalability of our proposed methods, we learn node representations on the Erdos-Renyi graphs using \ourwork-SP (default parameter settings, 10 thread). 
The graphs are generated with constant degree of 10 and sizes ranging from 100 to 1,000,000. 
We plot the running time \wrt the sizes of the graphs in Fig.~\ref{fig:scale} (in log-log scale).
The sampling procedure comprises of the role identification procedure and the random walk simulating procedure. The optimization procedure is the language model training with negative sampling.
From the figure, we can observe that \ourwork-WL scales super-linearly with the complexity of about $O(n^{1.2})$, which indicates that \ourwork is capable for massive networks.

\section{Conclusions}
\label{sec:conclusion}
We propose a new paradigm \ourwork for learning structural representations. 
\ourwork generates relabeled subgraphs by relabeling context nodes with their structural roles in the anchor node's local topologies, and then apply typical network embedding methods to learn structural embeddings. 
We develop two different relabeling methods.
Comprehensive experiments prove the effectiveness and efficiency of our proposed methods. 
The results of the experiments show that the learned structural embeddings are robust and can transfer across networks.
Our experiments also show differences between typical network embedding and structural embedding.

In the future, we plan to investigate a good way to combine typical network embeddings and structural embedings, such that we can capture both communiy memberships and structural roles.
Besides, we are also interested in investigating the potentials of structural embeddings in transfer learning tasks.

\section*{Acknowledgment}

This work was carried out with the support of the ``Fundamental Research Funds for the Central Universities, JLU'', ``Jilin Provincial Key Laboratory of Big Data Intelligent Computing (20180622002JC)'' and ``Jilin Natural Science Foundation (20180101036JC)''.

\bibliographystyle{IEEEtran}
\bibliography{NRL-NE2}

\end{document}